\documentclass[aps,pra,reprint,groupedaddress,amsmath,amssymb]{revtex4-2}
\usepackage{graphicx,xcolor}
\usepackage[colorlinks=true, linkcolor=blue, citecolor=blue, urlcolor=blue]{hyperref}
\usepackage{tabularx}
\usepackage{array}
\usepackage[T1]{fontenc}

\begin{document}
\newcommand{\be}{\begin{eqnarray}}
\newcommand{\ee}{\end{eqnarray}}
\newcommand{\pr}{\prime}
\newcommand{\bbm}{\begin{bmatrix}}
\newcommand{\ebm}{\end{bmatrix}}
\newcommand{\bfk}{\boldsymbol{k}}
\newcommand{\bfr}{\boldsymbol{r}}
\title{Exciton-polariton condensates in epsilon-near-zero cavities}

\author{Ege Özgün}
\email{egeozgun@hacettepe.edu.tr}
\affiliation{Department of Physics Engineering, Hacettepe University,
06800 Ankara, Türkiye}

\begin{abstract}
We propose an epsilon-near-zero (ENZ) artificial cavity that can enhance light-matter coupling around the ENZ frequency and theoretically show the possibility of exciton-polariton condensation in the suggested platform. By using a rate equation model based on the driven-dissipative reservoir approach, we calculate the pump threshold for condensation and then compare the suggested platform's advantages/disadvantages with distributed Bragg reflector cavities. We also discuss the validity of the mean field theory within our model.   
\end{abstract}


\maketitle

\section{Introduction}

Polariton condensates (PC)s offer higher critical temperature ($T_c$) values due to their lighter masses as compared to the atomic Bose-Einstein condensates \cite{BEC1,BEC2} and exciton condensates \cite{EC} allowing even room temperature $T_c$'s \cite{room_PC}. Most of the conventional PC experiments consist of distributed Bragg reflector (DBR) based microcavities. These can trap electromagnetic field modes in the form of standing waves due to high reflectance provided by the DBR layers on each side that are made up of layers with alternating dielectric values of thickness $\lambda/4$ each, to strongly contain the field with wavelength $\lambda$ in the microcavity region \cite{Atac}. Beyond conventional DBR microcavities, alternative polaritonic architectures based on Tamm-plasmon modes at metal–DBR interfaces were proposed \cite{Tamm-theory} and realized \cite{Tamm-exp}. Exciton polaritons have also been experimentally realized in semiconductor waveguide geometries \cite{waveguide}. Moreover, strong coupling between excitonic emitters and surface plasmon-polariton (SPP) modes localized at metal-dielectric interfaces were extensively studied \cite{plasmon-review}. Further details on PCs can be found in the comprehensive review papers \cite{Deng2010,Kavokin2010,Richard2010,Keeling2011,Carusotto2013,Byrnes2014,Keeling_R2020}.

Epsilon-near-zero (ENZ) materials \cite{Silveirinha2006} became popular by the turn of the millennium with their exceptional features including wavelength stretching, supercoupling and field enhancement \cite{Engheta2013}. There are different ways to achieve ENZ modes: Operating metal-clad waveguides near their cutoff frequencies, stacking layers of oxides (with positive permitivity) and metals (with negative permittivity) and more naturally thin metal films around their plasma frequencies can yield ENZ behavior. One utilization of ENZ materials is enhancing light-matter coupling \cite{Campione2015}, which is one of the essential components of obtaining polaritons.  

\begin{figure}[t]
\centering
\includegraphics[width=\columnwidth]{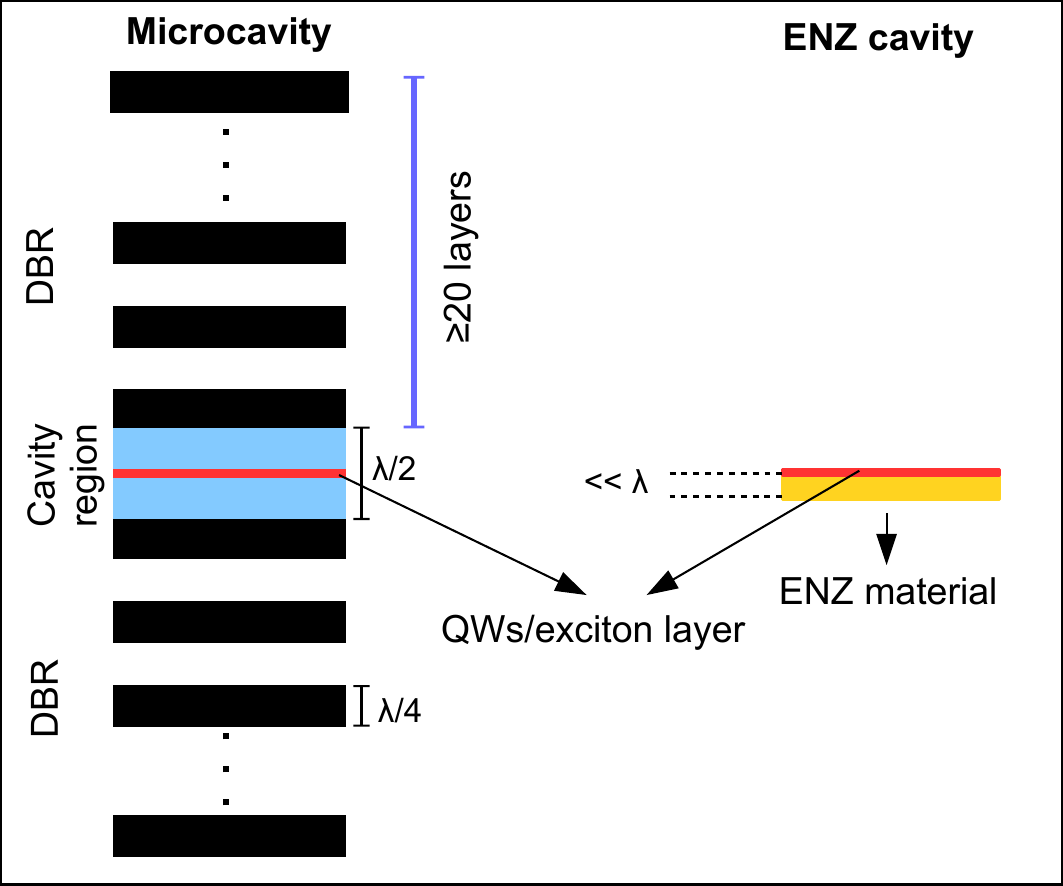}
\caption{Comparison of DBR microcavities and the suggested ENZ cavity.}
\label{dbr-enz}
\end{figure}

Here we theoretically study an alternative platform for achieving exciton PC (from now on we will simply call it PC) via ENZ artificial cavities. Fig. \ref{dbr-enz} shows the comparison between conventional DBR microcavities and the suggested ENZ cavity. In DBR cavities the real part of the dielectric function is greater than zero with ideally nearly zero losses ($\mathrm{Re}(\varepsilon)>0$, $\mathrm{Im}(\varepsilon)\approx 0$). To achieve high reflectance from the DBRs more than $20$ layers on each side are required \cite{Atac}, yielding a total size $> 10\lambda$ of the combined microcavity-exciton setup to trap the electromagnetic wave of wavelength $\lambda$. In the ENZ cavities, the real part of the dielectric function is ideally zero but as a downside there are considerable losses ($\mathrm{Re}(\varepsilon) \approx 0$, $0 < \mathrm{Im}(\varepsilon) < 1$). However the total size of the ENZ cavity-exciton structure $  \ll \lambda$ is at least an order of magnitude smaller than its DBR counterpart. Hence one advantage of the ENZ cavity would be the considerable reduction of the physical size. Another crucial difference is in the photonic content of the polaritons: In DBR cavities, the photonic part comes purely from the extended transverse standing wave mode trapped inside the cavity, in contrast for the ENZ cavities photonic part arises due to material assisted strongly confined longitudinal ENZ mode.

A comparison with SPP platforms is also crucial. In conventional SPP platforms, the electromagnetic mode is a surface-bound wave localized at a metal-dielectric interface, whereas the ENZ platform that we suggest exploits the longitudinal ENZ mode supported by the ultrathin metallic film near the zero crossing of its permittivity. The ENZ mode is characterized by a strongly enhanced electric-field component normal to the film, providing a distinct mechanism for light-matter coupling and polariton formation from that of conventional SPP-based systems.

A significant difference between the widely used planar cavities and the ENZ PC arises due to the unusual flat dispersion of the ENZ mode that can yield a small number of excitonic emitters per relevant photonic modes. When the number of excitonic emitters per relevant photonic modes is not large enough, validity of the mean field theory (MFT) becomes questionable. Therefore, it is important to study this ratio for the suggested ENZ platform.  

The remaining manuscript is structured in the following way: In Sec. \ref{2} we calculate the ENZ mode of the metal thin film, in Sec. \ref{3} we calculate polariton modes, in Sec. \ref{4} we study the pump threshold with two different models to find out which $k$ mode is expected to condense first, in Sec. \ref{5} we discuss the validity of the MFT, in Sec. \ref{6} we compare DBR and ENZ cavities within the context of PC and we conclude with Sec. \ref{7}.

\section{ENZ mode in Metal Thin Film}
\label{2}
We base our ENZ cavity on metal thin films. Depending on the plasma frequency of the material, when the thickness is reduced below a certain value one of the plasmon modes turns into an ENZ mode displaying flat dispersion over a wide k range \cite{campione2015theory}. We choose a 2 nm metal thin film with a plasma wavelength of 1 $\mu$m (for which the thicknesses below 10 nm yields ENZ modes) where above and below we assume the thin film is surrounded by air ($\varepsilon_1,\varepsilon_3=1$). Solving the Maxwell equations in three regions and imposing the boundary conditions yields the transcendental equation for calculating the modes \cite{campione2015theory}:

\be
1 + \frac{\varepsilon_1 k_{z3}}{\varepsilon_3 k_{z1}} = i~{\mathrm{tan}(k_{z2}d)} \left ( \frac{\varepsilon_2(\omega) k_{z3}}{\varepsilon_3 k_{z2}}+ \frac{\varepsilon_1 k_{z2}}{\varepsilon_2(\omega) k_{z1}} \right )
\label{enz_modes}
\ee

where $k_{zi}$ for $(i=1,2,3)$ is the longitudinal wavevector satisfying $k_{zi}^2+k_{\parallel}^2=\varepsilon_i(\omega/c)^2$. We take the normalized Drude decay $\gamma=10^{-2}$ (in units of $\omega_p$) and the permittivity for the thin metal film is given by $\varepsilon_2(\omega)=1-1/\left(\omega^2+i\gamma\omega\right)$ where the decay rate and all frequencies are normalized with the plasma frequency $\omega_p$. We further use the dimensionless wavevector $k=k_{\parallel}c/\omega_p$ throughout the manuscript.

\begin{figure}[h]
\centering
\includegraphics[width=\columnwidth]{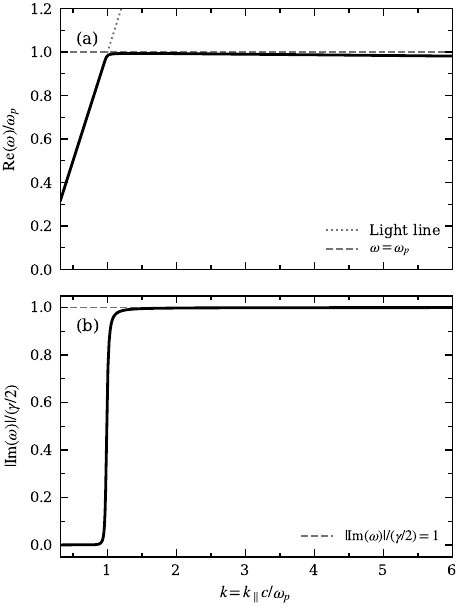}
\caption{(a) Dispersion of the ENZ mode and (b) the losses for the ultrathin metal film.}
\label{enz-disp}
\end{figure}

In the numerical calculations, a cut-off is chosen $(k=6)$ for the normalized in plane wavevector consistent with fabrication standards. The dispersion of the ENZ mode and the corresponding losses are given in Figure \ref{enz-disp}. The flat dispersion for the wide range of $k$ values is the strong signature of the ENZ mode and corresponding losses are the unavoidable outcome of the ENZ physics. Now the question is will the strong electric field confinement be sufficient to yield strong light-matter coupling interaction for obtaining polariton modes in the presence of these photonic losses. To answer this qusetion we calculate the polariton modes in the next section.

\section{Polariton modes}
\label{3}

We are now in a position to see if one can get strong light-matter coupling with this setting. A non-Hermitian model is used to include the losses from excitonic and photonic parts within the context of a Jaynes-Cummings type model in which the excitonic part is assumed to be uniform inside the thin film and taken to be dispersionless so photonic modes at different k couple to a single exciton mode. The effective $2\times2$ Hamiltonian we used is given below ($\hbar=1$):

\begin{figure}[h]
\centering
\includegraphics[width=\columnwidth]{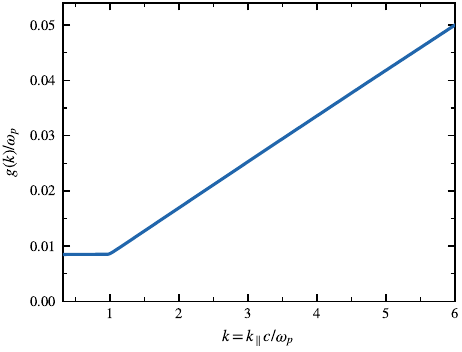}
\caption{Normalized light-matter coupling strength $g(k)/\omega_p$ obtained from the ENZ electric-field amplitude.}
\label{gk}
\end{figure}

\be
H_{\rm eff}(k)=
\begin{pmatrix}
\omega_c(k)- i\,\dfrac{\Gamma_c(k)}{2} & g(k)\\[6pt]
g(k) & \omega_x- i\,\dfrac{\Gamma_x}{2}
\end{pmatrix}.
\label{Heff}
\ee

in which $k=k_{\parallel}c/\omega_p$ is the normalized in-plane wavevector, $\omega_c(k)$ is the bare photon (ENZ mode) dispersion obtained numerically for the thinfilm, $\omega_x=0.98 \omega_p$ is the bare exciton resonance, $g(k)$ is the light–matter coupling, $\Gamma_c(k)$ denotes photonic losses for the ENZ mode again derived numerically from the ENZ thin film and $\Gamma_x=0.001 \omega_p$ is the excitonic loss. In our simple model, light matter coupling strength depends on the electric field amplitude in the following way:

\be
g(k)=g_0\,f_E(k),
\qquad
f_E(k)=\frac{|E_z(k)|}{|E_z(k_{E_{max}})|},
\label{coupling_str}
\ee

where $g_0=0.05 \omega_p$. That value corresponds to a Rabi splitting of about $124$ meV, which is experimentally accessible in SPP \cite{SPP-rabi} and ENZ architectures \cite{ENZ-rabi1, ENZ-rabi2}. Light-matter coupling strength is plotted in Fig. \ref{gk}, where the electric field strength is obtained from the metal thin film numerically. The eigenvalues of $H_{\rm eff}(k)$ are given by:

\be
\omega_{\pm}(k)
=\frac{\Omega_+}{2}
\pm \frac{1}{2}\sqrt{\Omega_-^2+4g(k)^2},
\label{polaritoneigs}
\ee

\begin{figure}[h]
\centering
\includegraphics[width=\columnwidth]{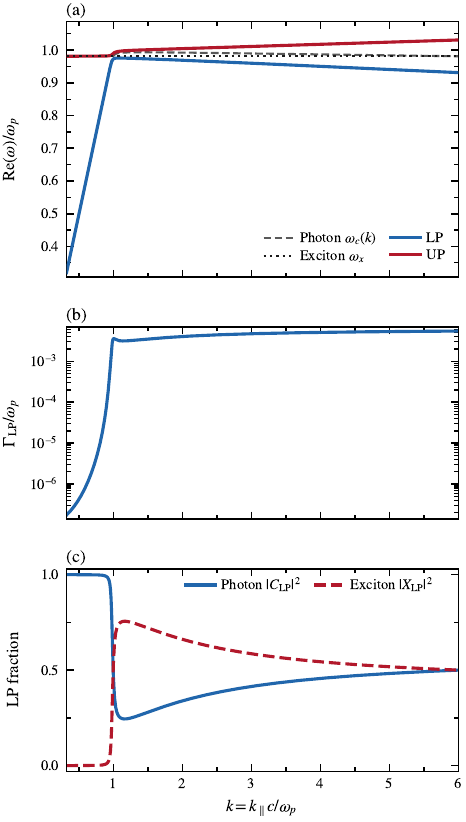}
\caption{(a) Polariton bands, (b) LP losses (log scale), (c) Excitonic and photonic content for LP.}
\label{polariton-comb}
\end{figure}

where, $\Omega_{\pm}=\Omega_c(k) \pm \Omega_x$, $\Omega_c(k)=\omega_c(k)- i \Gamma_c(k)/2$, $\Omega_x=\omega_x- i\Gamma_x/2$ and $+(-)$ corresponds to UP(LP). Since the resonant polariton splitting is $2g(k)$, an approximately linear momentum dependence of $g(k)$ is expected to produce a correspondingly linear variation of the splitting near resonance. However, the full separation between the polariton branches should be non-linear in general because the exciton-photon detuning is $k$-dependent. Fig. \ref{polariton-comb}a shows the polariton bands that display the expected behavior. Moreover the anticrossing at $k$ values corresponds to the onset of the ENZ regime. Fig. \ref{polariton-comb}b displays the losses for the LP branch in log scale where the lossy nature at ENZ region is clearly displayed. Fig. \ref{polariton-comb}c shows the Hopfield coefficients $C_{\mathrm{LP}}$ and $X_{\mathrm{LP}}$ describing photonic (ENZ mode) and excitonic content of the LP mode, respectively. The excitonic content is maximum around the $k$ values where the ENZ mode starts to grow. A peculiar difference in the obtained polariton modes for the ENZ cavity here as compared to the conventional DBR cavity polaritons is that the dispersion is not quadratic neither globally nor locally. This feature is due to the widely flat photonic dispersion brought by the ENZ mode.   

To calculate the emission spectrum, instead of using the non-Hermitian model, we start from a Jaynes-Cummings type Hamiltonian under rotating wave approximation in which each $k$ mode is treated independently:  

\be
\hat{H}(k)=\omega_{c}(k)\,\hat{a}_k^{\dagger}\hat{a}_k+\omega_{x}\,\hat{b}^{\dagger}\hat{b}+g(k) (\hat{a}_k^{\dagger} \hat{b} + \hat{b}^{\dagger} \hat{a}_k), 
\label{JC}
\ee

where $\hat{a}_k/ \hat{a}^{\dagger}_k$ annihilates/creates a photon mode $k$ and $\hat{b}/\hat{b}^{\dagger}$ is the excitonic annihilation/creation operator. Then we use the Lindblad master equation given below,

\be
\frac{d\hat{\rho}}{dt}
&=&-i\left[\hat{H}(k),\hat{\rho}\right]
+\Gamma_c(k)\,\mathcal{D}[\hat{a}_k]\,\hat{\rho}
\nonumber\\
&&+\Gamma_x\,\mathcal{D}[\hat{b}]\,\hat{\rho}
+P\,\mathcal{D}[\hat{b}^{\dagger}]\,\hat{\rho},
\label{lindblad}
\ee

\begin{figure}[h]
\centering
\includegraphics[width=\columnwidth]{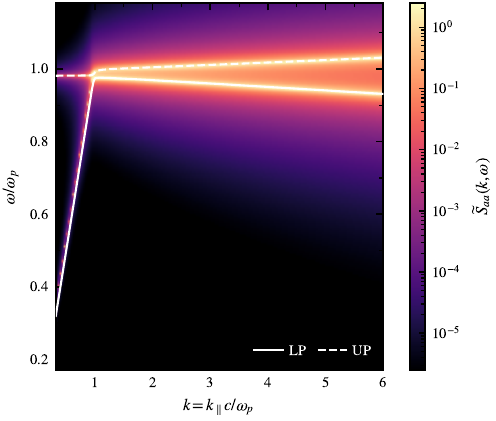}
\caption{Emission map for the polariton modes (in log scale) where $\widetilde{S}_{aa}(k,\omega) = \omega_p S_{aa}(k,\omega)$.}
\label{emission}
\end{figure}

with the Lindblad dissipator given by:

\be
\mathcal{D}[\hat{L}]\,\hat{\rho}=\hat{L}\hat{\rho}\hat{L}^{\dagger}-\frac{1}{2}\left(\hat{L}^{\dagger}\hat{L}\hat{\rho}+\hat{\rho}\hat{L}^{\dagger}\hat{L}\right).
\label{lindblad_dissipator}
\ee

where cavity loss, excitonic loss and incoherent pump collapse operators given by $\hat{L}_{c}(k)=\sqrt{\Gamma_c(k)}\,\hat{a}_k$, $\hat{L}_{x}=\sqrt{\Gamma_{x}}\,\hat{b}$ and  $\hat{L}_{P}=\sqrt{P}\,\hat{b}^{\dagger}$, respectively. Here $P$ is the weak exciton pump. Lindblad master equation is solved for each k value. After solving the Lindblad master equation, the emission spectrum can be calculated from:

\be
S_{aa}(k,\omega)=\frac{1}{\pi}\operatorname{Re}\int_{0}^{\infty}d\tau\,e^{i\omega\tau}\left\langle\hat{a}_{k}^{\dagger}(0)\hat{a}_{k}(\tau)\right\rangle_{\mathrm{ss}},
\label{emission_spectrum}
\ee

in which $\left\langle\hat{a}_{k}^{\dagger}(0)\hat{a}_{k}(\tau)\right \rangle_{\mathrm{ss}}$ is the non-equilibrium steady state two time correlator. The result for the UP and LP branches is given in Figure \ref{emission} which confirms sufficient strong light-matter coupling for the formation of well-defined UP/LP branches. 

\section{Pump threshold}
\label{4}

After confirming strong light-matter coupling, the following question is which k mode is expected to condense first and whether we can theoretically achieve PC for realistic values consistent with experiments. To test our expectations, we now calculate the pump threshold. To do this we adapt the driven-dissipative condensate model of Wouters and Carusotto \cite{wouters2007}. This model is normally written for a single condensate mode. Here we do this by retaining a single exact LP eigenmode at momentum $k_0$, neglect coupling to other modes, and take the reservoir-induced stimulated-scattering rate to be linear, $R(n_R)=R_{k_0}n_R$. This approach of considering a single condensate mode is equivalent to a mean field decomposition; we will study the validity of MFT in Section \ref{5}. The coupled condensate–reservoir mean-field equations are given by:

\begin{align}
\frac{dn_c}{dt}
={}&
\left[
R_{k_0}n_R
-\Gamma_{\mathrm{LP}}(k_0)
\right]n_c,
\nonumber\\
\frac{dn_R}{dt}
={}&
P-
\left(
\gamma_R+R_{k_0}n_c
\right)n_R ,
\label{density-rate-equations}
\end{align}

where $n_c$ is the condensate density (of mode $k_0$), $\Gamma_{\mathrm{LP}}(k_0)=|C_{\mathrm{LP}}(k_0)|^2\Gamma_c(k_0)+|X_{\mathrm{LP}}(k_0)|^2\Gamma_x$ gives LP losses, $n_R$ is the reservoir density, $\gamma_R$ gives reservoir decay rate, $R_{k_0}$ is the stimulated scattering into LP mode at $k_0$ and $P$ is the incoherent pumping rate. By linearizing Eq. \ref{density-rate-equations} (for the derivation see Appendix \ref{A}) we obtain the following expression for the pump threshold:

\begin{align}
P_{\mathrm{th}}(k_0)
={}&
\frac{\gamma_R}
{R_0|X_{\mathrm{LP}}(k_0)|^2}
\Big[
|C_{\mathrm{LP}}(k_0)|^2\Gamma_c(k_0)
\nonumber\\
&\qquad
+
|X_{\mathrm{LP}}(k_0)|^2\Gamma_x
\Big],
\label{pump-threshold}
\end{align}

where we expressed $R_{k_0}=R_0 \vert X_{\mathrm{LP}}(k_0)\vert^2$. We can now calculate the pump threshold $P_{\mathrm{th}}$ for all $k$ values and find out which mode is expected to go into condensation first.

To do that, we use two different models. In the first model, we use $\Gamma_{\mathrm{LP}}(k_0)$ without any modification. This approach will favor the low $k$ modes since before the ENZ mode onsets, in our model of the thin film the photonic losses are practically zero and since the excitonic loss is $k$-independent in our model, low $k$ values are expected to condense first. However this model is not realistic experimentally, thus we should incorporate the possible experimental losses to our model. So in the second model we define $\tau^{\ast}_{\mathrm{LP}}=10~{\text ps}$ as our maximum allowed LP lifetime \cite{Deng2010}, which corresponds to a minimum loss of $\Gamma_{min}=1/(\omega_p \tau^{\ast}_{\mathrm{LP}})$ and then redefine the LP losses with an effective model as $\Gamma^{\mathrm{eff}}_{\mathrm{LP}}(k_0)={\mathrm{max}}[\Gamma_{\mathrm{LP}}(k_0),\Gamma_{\mathrm{min}}]$. Thus, in the second model low $k$ modes below the ENZ region also has considerable photonic losses which is connected to the LP lifetime floor. In this model ENZ $k$ mode(s) are expected to condense first. For all calculations we use $\gamma_R=1/\tau_R=5.31 \times 10^{-6}~\omega_p$ with reservoir lifetime $\tau_R=100~{\mathrm{ps}}$ \cite{res_lif} and $R_0=2.65 \times 10^{-5}~\omega_p$ taken as a phenomenological scattering-rate scale. The corresponding reference time $R_0^{-1}=20~\mathrm{ps}$ lies within the $10$--$50~\mathrm{ps}$ range of LP-phonon relaxation times discussed in Ref.~\cite{Deng2010}.  

\begin{figure}[h]
\centering
\includegraphics[width=\columnwidth]{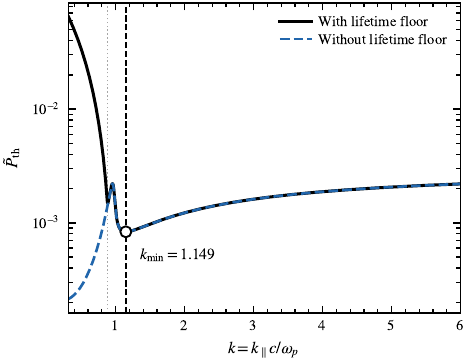}
\caption{Dimensionless pump threshold $\tilde{P}_{{\mathrm{th}}}=\tilde{\gamma}_R \tilde{\Gamma}_{{\mathrm{LP}}}/\tilde{R}$ where all rates normalized to $\omega_p$ with and without lifetime floor. Dotted and dashed lines show where two models become equal and the minimum $k$ value for the model with lifetime floor, respectively.}
\label{lifetime_floor}
\end{figure}

The results for the two models are compared in Fig. \ref{lifetime_floor}. When the lifetime floor is not included, the lowest $k$ value has the smallest $P_{{\mathrm{th}}}$ since photonic losses are practically zero before the ENZ region onsets and the excitonic loss is constant in our model. When we move deeper into the ENZ region, photonic losses increase substantially to the point where the lifetime floor no longer has an effect so in ${\mathrm{max}}[\Gamma_{\mathrm{LP}}(k_0),\Gamma_{\mathrm{min}}]$ the bare LP losses $\Gamma_{\mathrm{LP}}(k_0)$ is always dominant thus the two models start to yield the same result (dotted vertical line in Fig. \ref{lifetime_floor}). At $k=1.149$ there is a sweet spot between the excitonic content and the losses of the LP branch, yielding the $P_{{\mathrm{th}}}$ for the model with lifetime floor (black dashed vertical line in Fig. \ref{lifetime_floor}).        

As the result of the two different models we used, the most robust and experimentally realistic $k$ mode to go under condensation first is $k_0=1.149$. It is important to note that in our simple model we treated each mode separately, did not consider mode switching and scattering between LP states at different $k$ and assumed a spatially uniform reservoir. 

\section{Validity of the MFT}
\label{5}

While calculating the pump threshold for separate $k$ modes, we assumed MFT is valid. However the flat ENZ dispersion can yield large photonic DOS that can cause enhanced fluctuations, yielding the breakdown of the MFT approach. For a two-dimensional parabolic dispersion (which is the case for conventional DBR PCs), the density of states is constant. In that case one can directly check the validity of the MFT via a dimensionless photonic mass parameter as defined in Ref. \cite{keeling-MF1}. Since our photonic (ENZ) dispersion is far from being quadratic, we need to compare the amount of excitonic modes per photonic modes that contributes to the relevant energy window set by the interaction strength. This means we need to integrate the photonic DOS over that energy window defined by:

\be
\vert \omega_c(k)-\omega_c(k_0) \vert \leq g(k_0).
\label{energy_window}
\ee

We can now integrate the photonic DOS within that energy window to calculate the relevant number of photonic modes per area:

\be
\rho_{\rm ph}^{\rm rel}(k_0)
=
\int_{\omega_c(k_0)-g(k_0)}
^{\omega_c(k_0)+g(k_0)}
D_{\rm ph}(\omega)\,d\omega,
\label{rev_photonic}
\ee
where $D_{\rm ph}(\omega)$ is the photonic DOS given by:

\be
D_{\rm ph}(\omega)
= \left(\frac{\omega_p}{c}\right)^2
\int \frac{d\theta}{(2\pi)^2} k\,dk\,
\delta\!\left[\omega-\omega_c(k, \theta)\right],
\label{ph_DOS}
\ee

with $\delta$ denoting the Dirac delta function. Here we again used dimensionless parameters defined by $k=k_{\parallel}c/\omega_p$, $k_{\parallel} ~dk_{\parallel}=(\omega_p/c)^2 k~dk$. Inserting Eq. \ref{ph_DOS} in Eq. \ref{rev_photonic} and performing the azimuthal integral we have:

\be
\rho_{\rm ph}^{\rm rel}(k_0)
=
\frac{1}{2\pi}  \left(\frac{\omega_p}{c}\right)^2
\int_0^{k_{\max}}
k\,dk\,
\Theta\!\left[
g(k_0)
-
\left|\omega_c(k)-\omega_c(k_0)\right|
\right],
\nonumber \\
\label{ph_rel_final}
\ee

where $\Theta$ is the Heaviside step function. We can now use Eq. \ref{ph_rel_final} to calculate the relevant number of photonic modes per area at each $k_0$ value. We use four representative values in the range $\rho_{X}=10^{9}-10^{12}~\text{cm}^{-2}$ for the areal density of excitonic emitters \cite{exciton-density1, exciton-density2, exciton-density3}. Since we assumed a homogeneous planar system (metal thin film) in which excitons and photon modes occupy the same active lateral area, area terms cancel and the ratio $\rho_{X}/\rho_{\rm ph}^{\rm rel}(k_0)$ directly measures the number of excitonic states per relevant photon mode. 

\begin{figure}[h]
\centering
\includegraphics[width=\columnwidth]{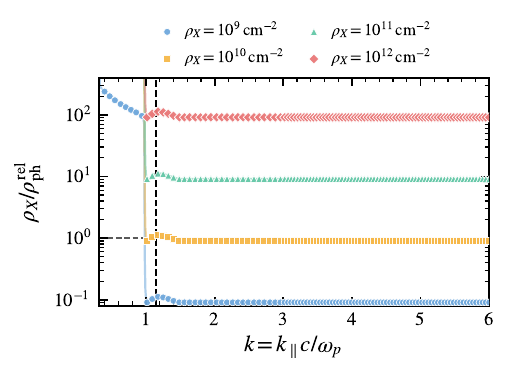}
\caption{Ratio of excitonic modes to relevant photonic modes for four different $\rho_X$ values. The condensing mode $k_0=1.149$ is shown with the vertical dashed line, whereas the horizontal dashed line shows when the ratio is unity.}
\label{MF_validity}
\end{figure}

Fig. \ref{MF_validity} displays $\rho_{X}/\rho_{\rm ph}^{\rm rel}(k_0)$ in which the dashed vertical line shows the condensing mode at $k_0 =1.149$. When $\rho_X/\rho_{\rm ph}^{\rm rel}\gg1$, many excitonic degrees of freedom couple to each relevant photon mode, suppressing the fluctuations and providing a regime in which the MFT is expected to be well defined. In contrast, when $\rho_X/\rho_{\rm ph}^{\rm rel}\sim1$, there is no large emitter-to-photon-mode ratio and fluctuation effects neglected within MFT can become important. At the condensing mode, our results give $\rho_X/\rho_{\rm ph}^{\rm rel}\simeq11$ and $111$ for $\rho_X=10^{11}$ and $10^{12}~\mathrm{cm}^{-2}$, respectively, hence these densities are an increasingly favorable regime for MFT. For $\rho_X=10^{10}~\mathrm{cm}^{-2}$, the ratio is close to unity, indicating a regime in which the validity of MFT is questionable. For $\rho_X=10^{9}~\mathrm{cm}^{-2}$, the relevant photon-mode density exceeds the excitonic emitter density, and the MFT is not expected to be valid. Thus, within the present mode-counting criterion, densities $\rho_X\gtrsim10^{11}~\mathrm{cm}^{-2}$ provide a comparatively safe regime for our single mode calculations within MFT. At lower densities, particularly when studying fluctuation-driven effects, competition and scattering between polariton modes, or mode switching, treatments beyond MFT may be required \cite{keeling-MF2}.
 
\section{DBR versus ENZ PCs}
\label{6}

After theoretically showing the possibility of ENZ PCs and validating our mean-field approach, a comparison with DBR PCs is needed. A striking difference is in the size of the structures, in which ENZ cavities can offer at least an order of magnitude size reduction as compared to DBR cavities. Second difference comes in the photonic content of the LPs: The conventional DBR PCs are commonly operated with a substantial photonic component and the electric field mode is due to extended transverse standing wave mode trapped inside the cavity. For the ENZ case, the photonic content is coming from the longitudinal ENZ mode and is strongly confined due to the ENZ nature of the material. Photonic/excitonic content in the LP mode has two sides: For the conventional DBR PCs in which the photonic dispersion is quadratic, larger photonic content generally reduces the polariton effective mass and thereby raises the characteristic temperature scale for condensation within MFT, whereas an LP with larger excitonic content can benefit from the incoherent pumping more since it feeds the excitonic reservoir. In our ENZ setting, the photonic dispersion is highly non-parabolic. Therefore we should not directly quote the same conclusion for the photonic content and be more cautious. Our MFT analysis in the previous section showed that for excitonic areal densities $\rho_X\gtrsim10^{11}~\mathrm{cm}^{-2}$ mean-field approach is safer to adapt, where as for lower densities the use of MFT is questionable.

For the suggested ENZ PC, we showed that the first $k$ mode to condense should be $k_0=1.149$ which has $\approx 75.5 \%$ excitonic and $\approx 24.5 \%$ ENZ mode content, which is mostly excitonic. Although the optical content is relatively small ENZ-mode component can still provide strong local optical overlap and can remain optically accessible with suitable outcoupling. Here we should discuss the optical access to the polariton modes in ENZ platforms. The conventional DBR PCs are commonly formed at small in-plane momentum and can radiatively couple through the cavity mirrors to freely propagating modes. However the ENZ mode considered here is a high-momentum confined mode lying outside the free-space light cone \cite{campione2015theory}. Thus, direct far-field coupling is prohibited by in-plane momentum conservation. Consequently one needs an additional momentum-matching mechanism, such as a diffraction grating or prism coupling in the Kretschmann configuration \cite{campione2015theory} for efficient optical outcoupling. The same considerations apply to direct resonant excitation of the ENZ mode, whereas non-resonant pumping of the excitonic reservoir we studied here does not require direct momentum matching to the condensate mode. Also the large excitonic content provides advantage in pumping that feeds the excitonic reservoir. That specific $k_0=1.149$ mode benefits from a favorable balance between LP loss and excitonic content yielding the minimum pump threshold $P_{{\mathrm{th}}}$ in our model.

\section{Conclusion}
\label{7}

Using coupled condensate-reservoir mean-field equations with k-dependent polariton parameters extracted from a non-Hermitian Jaynes-Cummings-type model, we theoretically demonstrated the possibility of exciton-polariton condensation in ENZ cavities. The underlying ENZ mode dispersion, losses, and electric-field amplitudes were obtained numerically for a thin metallic film and incorporated into the polariton model. The gain–loss balance of the incoherently pumped system identifies the lowest-threshold mode at $k_0=1.149$, which is predominantly excitonic. We showed that for excitonic areal densities $\rho_X\gtrsim10^{11}~\mathrm{cm}^{-2}$ MFT is expected to be valid. At lower densities, the reduced number of excitonic states per relevant photon modes suggests that fluctuation and multimode effects may become important. A quantitative description of this regime would require a treatment beyond the present mean-field model, including multimode dynamics and fluctuation effects, which we leave for future work. Our results demonstrate that ENZ modes can provide a viable platform for polariton condensation while offering at least an order-of-magnitude reduction in cavity dimensions compared with conventional DBR microcavities.

\begin{acknowledgments}
This study was supported by the Scientific and Technological Research Council of Turkey (TÜBİTAK) under project number 122F336 and by Hacettepe University Scientific Research Projects Coordination Unit under project number FHD-2025-22455. The author thanks Jonathan Keeling for insightful discussions.
\end{acknowledgments}

\makeatletter
\let\auto@bib@innerbib\@empty
\makeatother

\appendix
\section{Derivation of Eq. \ref{pump-threshold}}
\label{A}

For the second line of Eq. \ref{density-rate-equations}, the threshold is achieved when $n_c=0$ so,

\be
\frac{dn_R}{dt}= P-\gamma_R n_R.
\label{der4}
\ee
Reservoir stationary state is reached when $\frac{dn_R}{dt}=0$ so we have,

\be
n^{(0)}_R = \frac{P}{\gamma_R}.
\label{der5}
\ee
The condensate begins to grow when $R_{k_0} n^{(0)}_R = \Gamma_{\mathrm{LP}}(k_0)$ as it can be seen from the first line of Eq. \ref{density-rate-equations}, so the pump threshold is given by:

\be
P_{th} = \frac{\gamma_R \Gamma_{\mathrm{LP}}(k_0)}{R_{k_0}}.
\label{der6}
\ee
Inserting $R_{k_0}=R_0 \vert X_{\mathrm{LP}}(k_0)\vert^2$ and $\Gamma_{\mathrm{LP}}(k_0)=|C_{\mathrm{LP}}(k_0)|^2\Gamma_c(k_0)+|X_{\mathrm{LP}}(k_0)|^2\Gamma_x$ we have the final form of the expression:

\begin{align}
P_{\mathrm{th}}(k_0)
={}&
\frac{\gamma_R}
{R_0|X_{\mathrm{LP}}(k_0)|^2}
\Big[
|C_{\mathrm{LP}}(k_0)|^2\Gamma_c(k_0)
\nonumber\\
&\qquad
+
|X_{\mathrm{LP}}(k_0)|^2\Gamma_x
\Big],
\end{align}

\end{document}